\documentclass[a4paper,11pt]{article}

\usepackage{jcappub} 
\usepackage{lineno}

\usepackage{graphicx, amsmath,hyperref,natbib}
\usepackage[paperwidth=210mm,paperheight=297mm,centering,hmargin=1.8cm,vmargin=2.5cm]{geometry}
\usepackage{float}
\usepackage{amsfonts}
\usepackage{wasysym}
\graphicspath{{Bilder/}{../Bilder/}}
\usepackage{siunitx}
\newcommand{\qbounce}{{\it{q}}{\sc{Bounce}}}
\newcommand{\cannex}{{\sc{cannex}}}
\title{\boldmath Numerical Methods for Scalar Field Dark Energy in Table-top Experiments and Lunar Laser Ranging}

\author{Hauke Fischer}
\author{and Ren\'e I.P. Sedmik}

\affiliation{Atominstitut, Technische Universität Wien, Stadionallee 2, A-1020 Vienna, Austria}

\emailAdd{hauke.fischer@tuwien.ac.at, rene.sedmik@tuwien.ac.at}

\abstract{Numerous tabletop experiments have been dedicated to exploring the manifestations of screened scalar field dark energy, such as symmetron or chameleon fields. Precise theoretical predictions require simulating field configurations within the respective experiments. This paper focuses onto the less-explored environment-dependent dilaton field, which emerges in the strong coupling limit of string theory. Due to its exponential self-coupling, this field can exhibit significantly steeper slopes compared to symmetron and chameleon fields, and the equations of motion can be challenging to solve with standard machine precision. We present the first exact solution for the geometry of a vacuum region between two infinitely extended parallel plates. This solution serves as a benchmark for testing the accuracy of numerical solvers. By reparametrizing the model and transforming the equations of motion, we show how to make the model computable across the entire experimentally accessible parameter space. To simulate the dilaton field in one- and two-mirror geometries, as well as spherical configurations, we introduce a non-uniform finite difference method. Additionally, we provide an algorithm for solving the stationary Schr\"odinger equation for a fermion in one dimension in the presence of a dilaton field. The algorithms developed here are not limited to the dilaton field, but can be applied to similar scalar-tensor theories as well. We demonstrate such applications at hand of the chameleon and symmetron field. Our computational tools have practical applications in a variety of experimental contexts, including gravity resonance spectroscopy (\qbounce{}), Lunar Laser Ranging (LLR), and the upcoming Casimir and Non-Newtonian Force Experiment (\cannex{}). A Mathematica implementation of all algorithms is provided.}

\begin{document}

\maketitle
\flushbottom

\section{Introduction}

The quest to uncover the origins of dark energy stands as one of the most profound challenges in modern physics. Type Ia supernova observations have revealed that the Universe is currently expanding at an accelerated rate \cite{SupernovaCosmologyProject:1997zqe,SupernovaSearchTeam:1998fmf,SupernovaSearchTeam:1998bnz}, a finding supported by comprehensive cosmological studies \cite{Frieman:2008sn}. This accelerated expansion is mainly linked to a mysterious energy form referred to as dark energy. Two prominent strategies have been proposed to elucidate this cosmic acceleration. The first involves introducing a cosmological constant within the framework of general relativity, but this approach encounters intricate challenges related to fine-tuning \cite{Sola:2013gha} and is \emph{ad-hoc}, as it is not based on a fundamental understanding. An alternative approach proposes the introduction of a novel scalar field that interacts with gravity to explain this acceleration (see \cite{Joyce:2014kja} for a review). However, such scalar fields often give rise to so-called `fifth forces', which face stringent constraints imposed by tests of general relativity within the solar system. Consequently, these scalar fields require the inclusion of screening mechanisms to align with existing constraints. Several screening mechanisms, including the chameleon \cite{Khoury:2003rn,Khoury:2013tda}, K-mouflage \cite{Brax:2012jr,Brax:2014wla} and Damour-Polyakov \cite{Damour:1994zq} mechanisms, have been proposed. These mechanisms share a common feature of field suppression in dense environments resulting in the suppression of forces on large objects, as the field primarily couples only to a thin-shell near the object's surface.

High-precision tabletop experiments, conducted in a vacuum environment and on a much smaller scale than astrophysical phenomena, provide ideal tools for detecting screened scalar fields. In a vacuum, the field experiences less suppression compared to other environments, and the smaller experimental scale further weakens the thin-shell effect \cite{Burrage:2017qrf}. Among the most commonly studied screened scalar field models are the chameleon field \cite{Khoury:2003rn}, which screens by increasing its mass in dense environments, and the symmetron field \cite{Hinterbichler:2010es}, whose coupling to matter is proportional to its field value. In dense environments, symmetrons are in their symmetric phase, with a coupling to matter driven to zero, while in low-density environments, they are in a broken symmetry phase with an appreciable coupling to matter. A less explored model, motivated by string theory, is the environment-dependent dilaton \cite{Damour:1994zq,Brax:2010gi,Brax:2018iyo}, which employs a combination of increasing its mass and lowering its coupling to matter as screening mechanisms in dense environments. The parameters of the symmetron and chameleon models have already been constrained by various experiments, including atomic interferometry \cite{Burrage:2014oza,Hamilton:2015zga,Burrage:2015lya}, torsion balances~\cite{Upadhye:2012rc}, gravity resonance spectroscopy \cite{Brax:2017hna,Pitschmann:2020ejb,Cronenberg:2018qxf,Jenke:2020obe}, neutron interferometry \cite{Lemmel:2015kwa,Fischer:2023eww}, precision atomic measurements \cite{Brax:2022olf}, and more.

In a previous work \cite{Fischer:2023koa,Fischer:2023eww}, we presented the first experimental constraints of the environment-dependent dilaton theory but did not address the distinct numerical challenges associated with this model, such as higher precision requirements and steeper slopes compared to chameleon and symmetron fields, arising from the exponential self-coupling of this model. 

The main focus of this article is to demonstrate how to solve the stationary equations of motion for the environment-dependent dilaton field in experimental settings across the entire experimentally accessible parameter space. 

In Section \ref{2}, we provide the theoretical background of the investigated scalar field theories. An analytically exact solution of the environment-dependent dilaton field inside a vacuum region between two infinitely extended plates is derived in Section \ref{3}. Section \ref{4} illustrates the application of a novel parametrization for the dilaton model along with a refined transformation of the equation of motion. This approach addresses precision challenges unique to this model and ensures solvability at machine precision, even for demanding parameter sets. In Section \ref{5}, we present a non-uniform finite difference method for one and two mirror geometries, as well as spherical geometries. We furhter show how to resolve the extreme slopes of this field and demonstrate the applicability of the new method to other models. Section \ref{6} shows how to solve the one dimensional stationary Schr\"odinger equation in the presence of screened scalar fields. Applications to gravity resonance spectroscopy (\qbounce{}), Lunar Laser Ranging (LLR), and the Casimir and Non-Newtonian Force Experiment (\cannex{}) are discussed in Section \ref{7}.
We provide a publicly accessible Mathematica implementation the discussed methods~\cite{H.Fischer} allowing for arbitrary precision calculations.

\section{Theoretical Background}\label{2}

In this section, we explore the theoretical framework of scalar field theories in the Einstein frame representation of general relativity. We introduce specific scalar-tensor theories, including the chameleon, symmetron, and environment-dependent dilaton models.

\subsection{Scalar Field Theories in the Einstein Frame} \label{subsec:einstein-frame}

The scalar field theories under consideration are defined by their action in the Einstein frame, as detailed in \cite{Fujii2003}:

\begin{align}\label{eq:action}
S &= \int  d^4x\, \sqrt{-g} \left( -\frac{m_\text{pl}^2}{2}\,R + \frac{1}{2}\,\partial_\mu\phi\,\partial^\mu\phi - V(\phi) \right) \nonumber \\
&\quad+\int d^4x\,\sqrt{-\tilde g}\,\mathcal{L}_\text{SM} (\tilde g_{\mu \nu},\psi_i).
\end{align}

Here, $V(\phi)$ is the self-interaction potential of the scalar field $\phi$, $\tilde g_{\mu \nu}=A^2(\phi)\,g_{\mu \nu}$ is the Weyl-rescaled metric, $R$ is the Ricci scalar, and $m_{\text{pl}}$ is the reduced Planck mass. The Lagrangian density $\mathcal{L}_\text{SM}$ encompasses the Standard Model (SM) fields $\psi_i$. In the non-relativistic limit, the equations of motion for $\phi$ and the fifth force on a point particle with mass $m$ are given by \cite{pitschmann2023high}:

\begin{align}
    \Box \phi &= -V_{\text{eff}, \phi} (\phi; \rho)\>, \label{eqn} \\
    \vec{f}_{\phi} &= - \beta (\phi) \frac{m}{m_{\text{pl}}} \vec{\nabla}\phi , \label{force}
\end{align}

where $\rho$ is the matter density, $\beta(\phi)$ is the coupling function, and $V_{\text{eff}} (\phi; \rho)$ is the effective potential defined as:

\begin{align}
    \beta(\phi) &:= m_{\text{pl}} \frac{d\ \text{ln}A}{d\phi}, \\
    V_{\text{eff}} (\phi; \rho)\>&:= V(\phi)+\rho A(\phi).
\end{align}

\subsection{The Environment-Dependent Dilaton Model}

Table \ref{Table1} provides an overview of scalar field models, encompassing the chameleon, symmetron, dilaton, and their respective self-interaction potentials $V(\phi)$ and Weyl-rescaling factors $A(\phi)$ \cite{Burrage:2017qrf,Brax:2018iyo}.

\begin{table}[h]
\centering
\renewcommand{\arraystretch}{1.7}
\begin{tabular}{ccc}
\hline\hline
Model & $V(\phi)$ & $A(\phi)$ \\
\hline
Chameleon & $\frac{\Lambda^{n+4}}{\phi^n}$ &${\rm e}^{\phi / M_c}$ \\
Symmetron  & $-\frac{\mu^2}{2}\phi^2 + \frac{\lambda_S}{4}\phi^4$ & $1 + \frac{\phi^2}{2M^2}$ \\
Dilaton & $V_0\, {\rm e}^{-\lambda \phi /m_{\text{pl}}}$ &$1 + A_2 \frac{\phi^2}{2m_{\text{pl}}^2}$ \\
\hline\hline
\end{tabular}
\caption{Scalar field models with their respective self-interaction potentials $V(\phi)$ and Weyl-rescaling factors $A(\phi)$. Adapted from \cite{Burrage:2017qrf,Brax:2018iyo}.}
\label{Table1}
\end{table}

The primary focus of this article is the environment-dependent dilaton. It is parametrized by the energy density $V_0$, the dimensionless factor $\lambda$, and the dimensionless coupling parameter $A_2$. The exponential self-interaction potential is motivated by the string dilaton $\chi$ in the strong coupling limit, where $V(\chi) \rightarrow 0$ for $\chi \rightarrow \infty$. In Ref.~\cite{Damour:1994zq}, it was assumed that the Weyl-rescaling has a minimum at some large value $\chi_0$. In proximity to this minimum, the coupling would then be proportional to $(\chi-\chi_0)^2$. We obtain the corresponding model by defining $\phi:= \frac{m_{\text{pl}}}{\lambda}(\chi-\chi_0)$, as demonstrated in a comprehensive derivation in \cite{pitschmann2023high}. To neglect couplings to matter of higher order, we additionally demand:
\begin{align}
    A_2 \frac{\phi^2}{m_{\text{pl}}^2} \ll 1. \label{cutoff}
\end{align}
The full coupling to matter is then given by:
\begin{align}
    \beta = A_2\frac{\phi}{m_{\text{pl}}}.
\end{align}
The potential minimum of the field in a region with density $\rho_M$ is given by \cite{Brax:2022uyh}
\begin{align}
    \phi_{M} = \frac{m_{\text{pl}}}{\lambda} W\big( \frac{\lambda^2 V_0}{A_2 \rho_M} \big), \label{minimum}
\end{align}
 where the Lambert W-function $W$ is defined is defined as the inverse function of $x{\rm e}^x$.

\section{Two mirror dilaton solution} \label{3}

In this section, we derive the exact dilaton solution inside a vacuum region enclosed between two infinitely extended plates with density $\rho_M$, positioned at $z < -d$ and $z > d$. This geometric configuration closely resembles the one of the \cannex{} experiment, as detailed in Section \ref{6}. It is important to highlight that our choice to derive an exact solution for this specific geometry stems from the unavailability of exact solutions for the geometries relevant to the other experiments discussed in this paper.

It is worth noting that in earlier work \cite{Brax:2022uyh} an approximate two-mirror solution has been derived by linearizing the equations of motion. In contrast, our current derivation yields an exact solution, albeit under the assumption of $\rho_V=0$. This solution allows for a validation of the numerical methods discussed in Section \ref{4}-\ref{5} in the non-linear regime, setting it apart from the previously employed linear approximation.

The 1-dimensional equation of motion is obtained by rewriting Eq. (\ref{eqn}) as
\begin{align}
\frac{d^2\phi}{dz^2} = V_{\text{eff},\phi}(\phi;\rho) =-\frac{\lambda}{m_{\text{pl}}} V_0 {\rm e}^{-\lambda\phi/m_{\text{pl}}} + A_2\rho\frac{\phi}{m_{\text{pl}}^2}. \label{DEQ}
\end{align}
We now search for a solution for $\phi$ minimizing $V_\text{eff}$ such that asymptotically $\phi(z)\rightarrow \phi_M$ for $z \rightarrow \pm \infty$.

\subsection{Solution inside the vacuum region}

Upon multiplying Eq. (\ref{DEQ}) by $\phi'$ and integrating over $z$  between the plates, we arrive at the following relationship:
\begin{equation}
\frac{1}{2} \left(\frac{d\phi}{dz}\right)^2 = V_0 {\rm e}^{-\lambda\phi(z)/m_{\text{pl}}} - V_0 {\rm e}^{-\lambda\phi_0/m_{\text{pl}}}. 
\label{approx2}
\end{equation}

In the last expression, the choice of the integration constant ensures that $\phi'(0) = 0$, a condition dictated by the symmetry of the experimental setup. We also defined $\phi_0$ as $\phi(0)$.

Subsequently, we introduce $u(z)$ through the relationship

\begin{equation}
\phi(z) = \phi_0 - \frac{m_{\text{pl}}}{\lambda} \ln \big(u(z)\big),
\label{slope}
\end{equation}
which implies that $u(0) = 1$. This leads to the equation
\begin{align}
\frac{m_{\text{pl}}^2}{\lambda^2} \frac{(u'(z))^2}{2u(z)^2} = V_0 {\rm e}^{- \lambda \phi_0 /m_{\text{pl}}} (u(z)-1). \label{C1a}
\end{align}
Defining further
\begin{align}
    \alpha := \sqrt{2 V_0} \frac{\lambda}{m_{\text{pl}}} {\rm e}^{-\lambda\phi_0/(2 m_{\text{pl}})}, \label{alpha}
\end{align}
allows us to rewrite Eq. (\ref{C1a}) as:
\begin{align}
\frac{u'(z)^2}{u^2(z)(u(z)-1)} = \alpha^2. \label{C2a}
\end{align}

We continue by solving Eq. (\ref{C2a}) in the region $-d \leq z \leq 0$. Within this region, $\phi$ is increasing toward its local maximum at $z=0$, consequently leading to $\phi'(z) \geq 0$. This, as indicated by Eq. (\ref{slope}), implies that $u'(z) \leq 0$. Therefore, integration results in:
\begin{align}
 \int_{z}^{0}\frac{u'(s)}{\sqrt{u^2(s)(u(s)-1)}}ds &= \alpha z.
\end{align}

By introducing the variable $y(s) := \sqrt{u(s) - 1}$, we obtain:
\begin{align}
2 \int_{\sqrt{u(z)-1}}^{0}\frac{1}{1+y^2}dy = \alpha z.
\end{align}
 This last expression can be readily solved for $u(z)$, resulting in:

\begin{align}
u(z) = 1 + \text{tan}\left(\frac{\alpha}{2}z\right)^2. \label{even}
\end{align}

 Through analogous reasoning, we can conclude that the expression found for $u(z)$  also solves Eq. (\ref{C2a}) in the region $0 \leq z \leq d$. Therefore, the complete solution between the plates can be expressed as:

\begin{align}
\phi(z) = \phi_0 - \frac{m_{\text{pl}}}{\lambda}\text{ln}\left[1 + \text{tan}\left(\frac{\alpha}{2}z\right)^2\right].\label{final}
\end{align}

\subsection{Boundary conditions and definition of $\phi_0$}

The discovered solution contains the free parameter $\phi_0$, which needs to be determined from boundary conditions. Multiplying Eq. (\ref{DEQ}) by $\phi'$ within the mirrors and integrating leads to:

\begin{align}
\frac{1}{2}\left(\frac{d\phi}{dz}\right)^2 &= V_{\text{eff}}(\phi; \rho_M) - V_{\text{eff}}(\phi_{M}; \rho_M). \label{Mirror}
\end{align}
Here, the integration constant must be chosen such that the derivative vanishes asymptotically for $z \rightarrow \pm \infty$. Demanding continuity of the derivative at $z=d$ results in an equation for $\phi_d := \phi(d)$. Equating (\ref{Mirror}) and (\ref{approx2}) at $z=d$ yields

\begin{align}
\phi_d = \sqrt{\frac{2 m_{\text{pl}}^2}{A_2 \rho_M}(V_{\text{eff}}\left(\phi_{M}, \rho_M) - V_0 {\rm e}^{-\lambda \phi_0/m_{\text{pl}}}\right)}, \label{bounday}
\end{align}
where we utilized the positivity of the dilaton field. Equating  (\ref{bounday}) with (\ref{final}) at $z=d$ finally results in an implicit equation that defines $\phi_0$:

\begin{align}
&\phi_0 - \frac{m_{\text{pl}}}{\lambda}\text{ln}\left[1 + \tan\left(\frac{\alpha}{2}d\right)^2\right]\notag \\
&= \sqrt{\frac{2 m_{\text{pl}}^2}{A_2 \rho_M}\left(V_{\text{eff}}(\phi_{M}, \rho_M) - V_0 {\rm e}^{-\lambda \phi_0/m_{\text{pl}}}\right)}\,.\label{realPhi0}
\end{align}

Due to the periodicity of the tan, multiple solutions for $\phi_0$ can exist. 
However, since $\phi(z)$ must be defined on the domain $-d \leq z \leq d$ and $\tan(x)$ is undefined for  $x = \pi/2 + n \pi$ and $\ n\in \mathbb{Z}$, the correct solution must satisfy $\alpha d < \pi$, establishing a lower bound on $\phi_0$:

\begin{align}
    \phi_0 > \frac{2m_{\text{pl}}}{\lambda}\text{ln}\left( \frac{\lambda \sqrt{2 V_0}d}{\pi m_{\text{pl}}}\right).
\end{align}

For the dilaton parameters we investigated, only one solution compatible with this lower bound was found. This solution will be instrumental in demonstrating the accuracy of the proposed methods for solving the equations of motion in Section \ref{4} and \ref{5}. Notably, when $\phi_0$ does not closely approach $\phi_V$, this solution serves as an excellent approximation for the more realistic scenario where $\rho_V>$ 0. \\

This observation becomes apparent when examining the region near the potential minimum $\phi_\rho$ defined by $V_{\text{eff},\phi}(\phi_\rho;\rho)=0$. Near $\phi_\rho$, the following approximation, deduced from Eq. \ref{DEQ}, holds:

\begin{align}
\frac{\lambda}{m_{\text{pl}}} V_0 {\rm e}^{-\lambda\phi/m_{\text{pl}}} \simeq A_2\rho\frac{\phi}{m_{\text{pl}}^2}. \label{zero}
\end{align}

Deep within the mirror, the field is suppressed to $\phi_M$, leading both terms on the right-hand side (RHS) of Eq. (\ref{DEQ}) to contribute roughly equally to the differential equation. As the field transitions into the vacuum region, the linear term undergoes a sudden suppression by several orders of magnitude due to $\rho_V \ll \rho_M$. Consequently, the exponential term begins to dominate Eq. (\ref{DEQ}). The field advances toward its new potential minimum, and it is only in proximity to $\phi_V$ that the linear term ceases to be suppressed, as indicated by Eq.~(\ref{zero}). Therefore, as long as $\phi_0 \ll \phi_V$, the linear term can be disregarded within the vacuum region, which is equivalent to setting $\rho_V=0.$ 

\section{Precision Challenges in Dilaton Field Computations for Table-Top Experiments} \label{4}

This section focuses on precision problems and their solution of the environment-dependent dilaton model in the experimentally relevant parameter space shown in Figure.~\ref{fig:PS}. 

\subsection{Phenomenologically relevant parameter space}

\begin{figure}[H]
\begin{center}
\includegraphics[width=1\linewidth]
{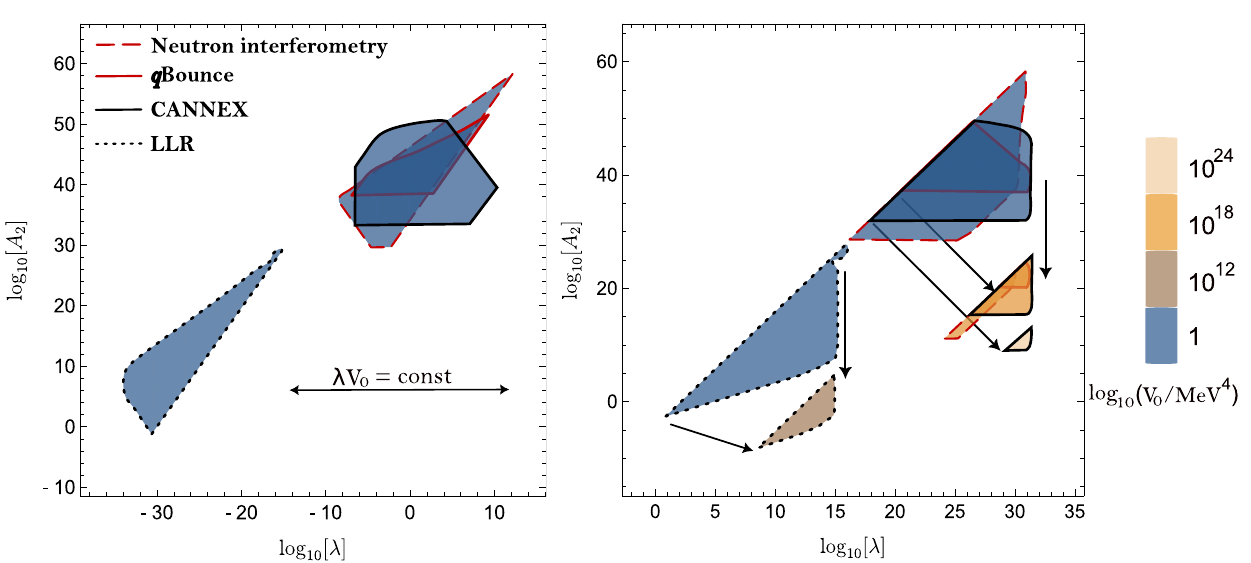}
\caption{Current and prospective constraints on the parameters of the environment-dependent dilaton are shown in colour~\cite{Fischer:2023koa}. The arrows show how the constraint areas shift for larger values of $V_0$. For small values of $\lambda$ (left), the physics only depends on the product $V_0\lambda$, but not on the individual values of $V_0$ and $\lambda$. Hence, the constraint areas simply shift towards smaller $\lambda$ for increasing $V_0$, without changing their shape. \protect\footnotemark}
\label{fig:PS}
\end{center}
\end{figure} 

\footnotetext{The presented limits illustrate the 'fermi-screening' approximation, wherein the neutron is treated as a sphere with a radius of 0.5 fm, aligning with Quantum Chromodynamics (QCD). In a separate study~\cite{Fischer:2023koa}, a secondary approximation known as 'micron-screening' was employed. This involved deriving a radius from the vertical extent of the wave function in the $\qbounce{}$ experiment.
Due to the distinct shapes of the neutron wave functions in neutron interferometry and the $\qbounce{}$ experiment, direct comparisons yield less meaningful results. It is essential to recognize that the wave functions in both cases do not align well with a spherical approximation. Consequently, we have chosen to exclusively present the more comparable and reliable constraints derived from the 'fermi-screening' approach.
It is noteworthy that these 'fermi-screening' constraints are more conservative, meaning they are encompassed within the constraints derived from the 'micron-screening' approximation.}

 The parameter space of the model can be broadly categorized into three regimes \cite{Fischer:2023koa}:
\begin{itemize}
    \item For small $\lambda$ (Figure~\ref{fig:PS} left), the approximation
    \begin{align}
        {\rm e}^{-\lambda \phi / m_{\text{pl}}} &\simeq 1
    \end{align}
    holds. Consequently, the equations of motion can be simplified to:
    \begin{align}
        \Box \phi = \frac{\lambda}{m_{\text{pl}}}V_0{\rm e}^{-\lambda \phi / m_{\text{pl}}} - A_2 \rho\frac{\phi}{m_{\text{pl}}^2} \simeq \frac{\lambda}{m_{\text{pl}}}V_0 - A_2 \rho\frac{\phi}{m_{\text{pl}}^2}.
    \end{align}  
   This results in linear equations of motion that can be solved exactly for various geometries. The parameter region characterized by small $\lambda$ was observed to be numerically straightforward and, therefore, is not explored in this article.    
    \item For intermediate values of $\lambda$, the model is undefined due to the violation of condition~(\ref{cutoff}).    
    \item For large $\lambda$ (Figure~\ref{fig:PS} right), the dilaton field exhibits strong exponential self-coupling, resulting in highly non-linear equations of motion where
    \begin{align}
        {\rm e}^{-\lambda \phi / m_{\text{pl}}} \ll 1.
    \end{align}
\end{itemize}
The parameter $V_0$ can take values as large as $V_0=10^{10^{24}}$ MeV$^4$ in the phenomenologically relevant parameter region, posing significant numerical challenges, which we elaborate on in the next subsection.

\subsection{Computing functions for very large values of $V_0$}

Large values of $V_0$ pose challenges such as overflows (numbers are too large to be represented on the computer)  and underflows (numbers are too small to be represented on a computer). For instance, the product 
\begin{align}
    V_0 {\rm e}^{-\lambda \phi / m_{\text{pl}}} \label{before}
\end{align}
can be physically sensible, but the individual values $V_0$ and  ${\rm e}^{-\lambda \phi / m_{\text{pl}}}$ may become extremely large or small,
leading to underflows in the exponential. To address this, we introduce the parameter $\gamma:= \text{log}_{10}(V_0/\text{MeV}^4)$, allowing us to rewrite the effective potential as

\begin{align}
    {\rm e}^{\big(-\lambda \phi / m_{\text{pl}}+ \gamma \text{ln}(10)\big)}\text{MeV}^4 + \frac{A_2 \rho}{2 m_{\text{pl}}^2}\phi^2,
\end{align}
and avoid underflows. Functions involving $\gamma$ need to be rewritten to avoid internal overflows. For example, evaluating the potential minimum~(\ref{minimum}) 

\begin{align}
    \phi_\rho = \frac{m_{\text{pl}}}{\lambda} W\left(\frac{\lambda^2 V_0}{A_2 \rho}\right) = \frac{m_{\text{pl}}}{\lambda} W\left(\frac{\lambda^2 {\rm e}^{\gamma \text{ln}(10)}\text{MeV}^4}{A_2 \rho}\right) \label{MIN}
\end{align}
can lead to overflow in ${\rm e}^{\gamma \text{ln}(10)}$. However, an asymptotic expansion of Lambert’s $W$ function allows us to use \cite{Corless:1996zz}
\begin{align}
W(x) &= \text{log }x - \text{ln }\text{ln}x + \sum_{k=0}^{\infty} \sum_{m=1}^{\infty}c_{km} (\text{ln }\text{ln}x)^m(\text{ln }x)^{-k-m}, \\
c_{km}&= \frac{1}{m!}(-1)^k \begin{bmatrix}k+m\\{k+1}\end{bmatrix},
\end{align}
where $\begin{bmatrix}k+m\\{k+1}\end{bmatrix}$ denotes the Stirling cycle number of first kind. Cancellation of $\text{ln}({\rm e}^{\gamma \text{ln}(10)})=\gamma \text{ln}(10)$ results to lowest order in
\begin{align}
    \phi_\rho \simeq \frac{m_{\text{pl}}}{\lambda}\Big( \gamma \text{ln}(10) + \text{ln}(\frac{\lambda^2 \text{MeV}^4}{A_2 \rho})\Big), \label{minapprox}
\end{align}
but higher order correction terms of $W$ can be added analogously. 

The final numerical challenge encountered for large values of $\gamma$ lies in the limitation of machine precision calculations, typically accurate up to 15-16 digits on present 64 bit machines. Such precision proves inadequate for deriving physically meaningful results. Consider a static field in a vacuum chamber with density $\rho_V$, surrounded by material walls with density $\rho_M$. The field is governed by the inequality 
\begin{align}
    \phi_M\leq \phi(x)\leq \phi_V.
\end{align}
For the maximum value of $\gamma \sim 10^{24}$ in the context of table-top experiments, the relative difference between $\phi_V$ and $\phi_M$ is approximately expressed as 
\begin{align}
   2 \frac{\phi_V-\phi_M}{\phi_V+\phi_M} \simeq \frac{\text{ln}(\frac{\rho_M}{\rho_V})}{\gamma \text{ln}(10)} =10^{-23}.
\end{align}
Here, we assume that in experimental setups, $\rho_M$ is roughly 10 orders of magnitude larger than $\rho_V$. Consequently, $\phi(x)=\phi_M=\phi_V$ to machine precision, providing no meaningful physical information. Hence, solving the equations of motion directly requires significantly higher precision which most software cannot provide. This higher precision also comes at a higher computational cost. 
In the subsequent subsection, we therefore introduce a method to circumvent calculations beyond machine precision.

\subsection{Circumventing the need for high precision calculations}

Machine precision calculations suffice for $\gamma$ values up to $10^{12}$. However, for $\gamma$ values exceeding this threshold, higher precision becomes imperative. In this section, we show how to circumvent this problem. This involves a reformulation of the dilaton field, expressed as:

\begin{align}
    \phi(x)=\phi_M+\delta(x). \label{newdef}
\end{align}
Here, $|\delta(x)|\ll|\phi_M|$ denotes a small value predominantly influencing only the less significant digits of $\phi$ for large $\gamma$ but significantly influencing the the dilaton force $\vec f_{\phi}$. Accordingly, we write:

\begin{align}
    \vec f_{\phi} = -\beta(\phi) \frac{m}{m_{\text{pl}}} \vec{\nabla} \phi = -\beta(\phi) \frac{m}{m_{\text{pl}}} \vec{\nabla} \delta.
\end{align}
Failing to single out $\delta$ and computing $\phi$ directly with machine precision would lead to the incorrect result $\phi(x)=\phi_M$, and hence $\vec{\nabla} \phi$ = 0. 
We proceed by deriving a differential equation which can be used to directly compute $\delta$. Starting from the dilaton field equation
\begin{align}
    \Box \phi  =   \frac{\lambda V_0}{m_{\text{pl}}} {\rm e}^{- \lambda \phi/m_{\text{pl}}} - \frac{A_2 \rho}{m_{\text{pl}}^2}\phi, \label{Original}
\end{align}
and inserting Eq.~(\ref{newdef}), we obtain
\begin{align}
    \Box \delta  =   \frac{\lambda V_0}{m_{\text{pl}}} {\rm e}^{- \lambda (\phi_M+\delta)/m_{\text{pl}}} - \frac{A_2 \rho}{m_{\text{pl}}^2}(\phi_M+\delta). \label{NewEQ}
\end{align}
In order to accurately compute $\delta$ from Eq. (\ref{NewEQ}) with machine precision, we need to carefully evaluate both terms on the right-hand side (RHS) in order not to introduce round-off errors. The term linear in $\phi$ is hardly influenced, as $\delta$ only affects the least significant digits. Rounding in the exponent of the first term, on the other hand, can cause serious numerical errors. This problem can be solved by using the multiplicity of the exponential and

\begin{align}
    V_{\text{eff},\phi}(\phi_M;\rho_M)=0 \Leftrightarrow \frac{\lambda V_0}{m_{\text{pl}}} {\rm e}^{- \lambda \phi_M/m_{\text{pl}}} = \frac{\beta(\phi_M)\rho_M}{m_{\text{pl}}}.
\end{align}
This results in: 

\begin{align}
    \Box \delta  =  \frac{\beta(\phi_M)\rho_M}{m_{\text{pl}}}{\rm e}^{-\lambda \delta / m_{\text{pl}}}- \frac{\beta(\phi_M+\delta)\rho}{m_{\text{pl}}}. \label{GoodEq}
\end{align}
This equation can be solved with machine precision over the entire experimentally relevant parameter space. In section \ref{DilatonTests} we validate the reliability of this approach for $\gamma=1$ and $\gamma = 10^{24}$. Subsequently, we focus on numerically solving the static equation of motion for the dilaton and similar scalar fields.

\section{Numerical solution of stationary scalar field equations of motion} \label{5}

Computing scalar fields in static configurations involves solving the partial differential equation:

\begin{align}
\frac{d^2\phi}{dx^2}+ \frac{d^2\phi}{dy^2}+\frac{d^2\phi}{dz^2} = V_{\text{eff},\phi}(\phi;\rho). \label{eq:phi_diffeq_explicit}
\end{align}

We will solve Eq.~(\ref{eq:phi_diffeq_explicit}) explicitly for the geometry of one and two mirrors, as well as a sphere, since these play a crucial role in the tabletop experiments discussed in Section \ref{6}. Additionally, we will briefly touch upon the generalization of these results to other geometries. The relevant geometries can be encapsulated by the following differential equation:
\begin{align}
\frac{d^2\phi}{ds^2}+ a\frac{2}{s}\frac{d \phi}{ds} = V_{\text{eff},\phi}(\phi;\rho)  \label{DEQ2}, 
\end{align}
where $s\in \{r,z\}$ for the spherical or one and two mirror geometries respectively. We introduced the parameter $a$ that takes the value 0 for one and two mirrors and the value 1 for a sphere. For one mirror, we assume a mirror with density $\rho_M$ at $s<0$ and a vacuum region with density $\rho_V$ above. Similarly, the sphere with radius $R$ has a density $\rho_M$ and is surrounded by vaccum with density $\rho_V$. The primary challenge in solving  Eq. (\ref{DEQ2}) is given by the extreme slopes of the dilaton field, as illustrated in Figure (\ref{fig:Comp1}). The dilaton field can show significant variations at scales as small as 10 fm for experimentally accessible parameters. An accurate uniform discretization would require up to $10^8$ grid points in the vacuum region alone, which is computationally unfeasible.

\subsection{Non-uniform finite difference method}

To address the issue of steep slopes, the utilization of a non-uniform mesh within the simulation interval becomes imperative. While the finite element method (FEM) (see e.g. \cite{langtangen2019introduction}) is a prevalent choice, allowing for an arbitrary mesh, this paper leverages non-uniform finite difference methods (FDM) due to their simpler implementation, particularly advantageous for the straightforward geometries considered. 

Starting from the grid points $s_0, ..., s_{N+1}$ of the simulation interval, we define
\begin{align}
    h_i &:= s_{i+1}-s_i \\
    \phi_i &:= \phi(s_i).
\end{align}
Taylor expanding to second order results in
\begin{align}
\phi_{i+1} &= \phi_{i} + \phi_{i}'h_i + \frac{\phi_{i}''}{2}h_i^2 + \mathcal{O}(h_i^3)\\
\phi_{i-1} &= \phi_{i} + \phi_{i}'h_{i-1} + \frac{\phi_{i}''}{2}h_{i-1}^2 + \mathcal{O}(h_{i-1}^3) .
 \end{align}
Neglecting higher order terms, we define the approximations of $\phi_i'$ and $\phi_i''$ as solutions to
$$
     \begin{pmatrix}
         h_i & \frac{h_i^2}{2} \\ 
         -h_{i-1} & \frac{h_{i-1}^2}{2}
     \end{pmatrix}
     \times
     \begin{pmatrix}
         \phi_{i}' \\ 
         \phi_{i}''  
     \end{pmatrix}
      \approx
     \begin{pmatrix}
         \phi_{i+1}-\phi_i \\ 
         \phi_{i-1} -\phi_i
     \end{pmatrix},
$$
 which leads to 

\begin{align}
\phi_{i}'' &\approx \frac{2(\phi_{i+1}-\phi_{i})}{h_i(h_i+h_{i-1})}-\frac{2(\phi_{i}-\phi_{i-1})}{h_{i-1}(h_i+h_{i-1})}, \label{dx} \\
\phi_{i}' &\approx \frac{h_i(\phi_i-\phi_{i-1})}{h_{i-1}(h_i+h_{i-1})}+\frac{h_{i-1}(\phi_{i+1}-\phi_{i})}{h_{i}(h_i+h_{i-1})}. \label{approx}
\end{align}

These approximations are an extension of the standard central difference scheme to non-uniform grids \cite{leveque2007finite} and maintain second order accuracy if $h_i = h_{i-1}$. Plugging Eqs.~(\ref{dx})--(\ref{approx}) into Eq. (\ref{DEQ2}) results in the discretized differential equation

\begin{align}
\frac{2(\phi_{i+1}-\phi_{i})}{h_i(h_i+h_{i-1})}-\frac{2(\phi_{i}-\phi_{i-1})}{h_{i-1}(h_i+h_{i-1})}+ a\frac{2}{s_i}\big[\frac{h_i(\phi_i-\phi_{i-1})}{h_{i-1}(h_i+h_{i-1})}+\frac{h_{i-1}(\phi_{i+1}-\phi_{i})}{h_{i}(h_i+h_{i-1})}\big] - V_{\text{eff}, \phi}(\phi_i, \rho_i) = 0. \label{DiscreteEq}
\end{align}

A method to solve this system of equations and apply boundary conditions is provided in Appendix \ref{AppSM1}. For geometries with less symmetry, Eq.~(\ref{dx}) has to be applied to each partial derivative separately. The main task is to choose $s_0,...,s_{N+1}$ appropriately to accurately capture the behavior of the scalar fields under consideration, as is discussed next. 

\subsection{Mesh construction}\label{MeshCon}

\begin{figure}[!ht]
\begin{center}
\includegraphics[scale=0.28]{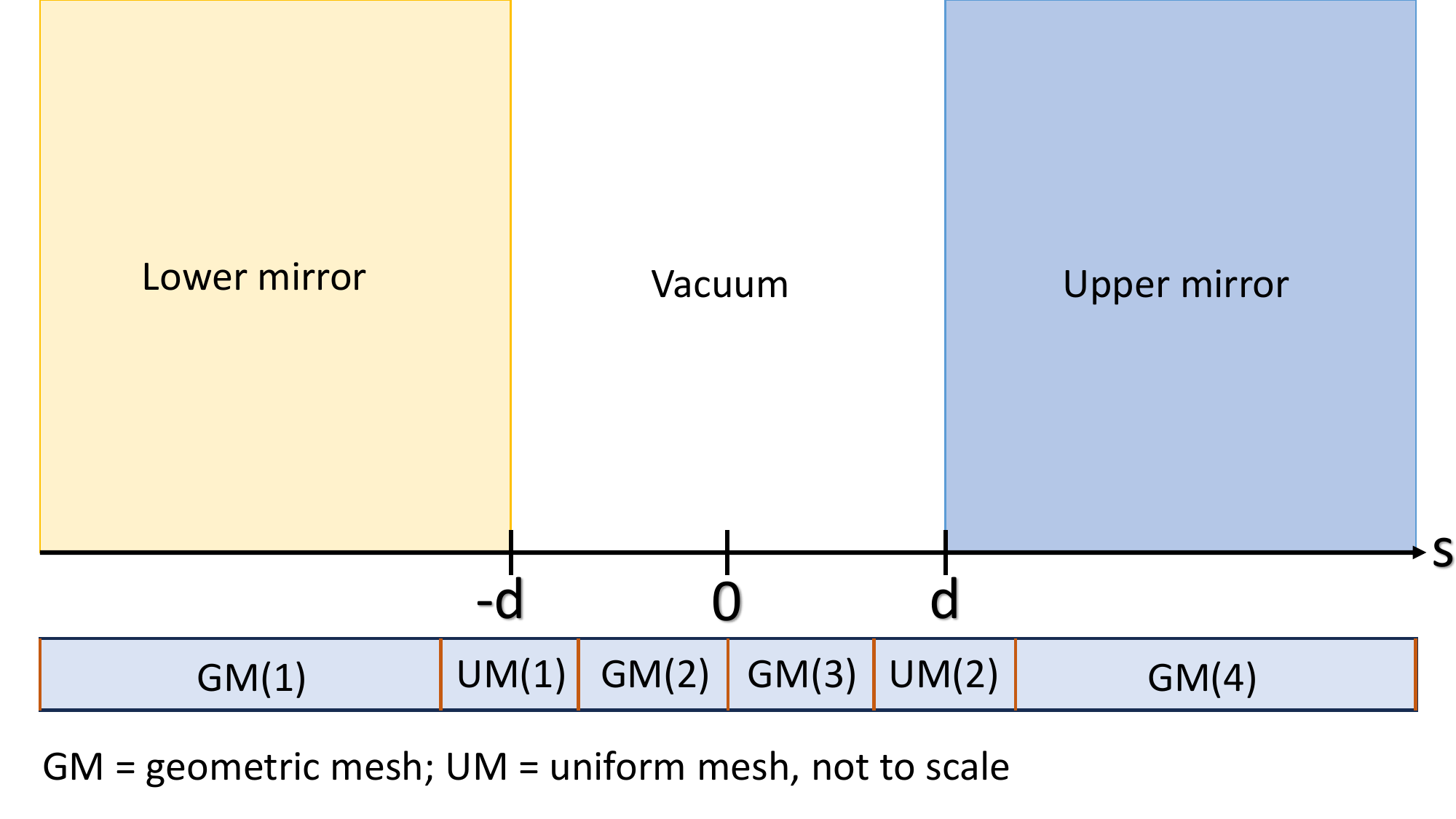}
\caption{Sketch for mesh construction, not to scale.}
\label{fig:mesh}
\end{center}
\end{figure} 

The screening mechanisms inherently suppress the scalar field within the mirrors, swiftly converging to their vacuum expectation value, $\phi_V$, within the vacuum region (for most parameter sets). To accurately represent the steep gradients near the transition from a mirror to the vacuum region, while efficiently discretizing regions of homogeneous densities, we employ a hybrid approach utilizing both uniform and geometric meshes \cite{brune2011exponential}, as illustrated in Figure \ref{fig:mesh} for a two mirror geometry. This approach requires only a few parameterizations.

\subsubsection{Uniform meshes:}
Around the material surfaces $s=-d$ and $s= d$ (designated as {\textsf{UM(1)}} and {\textsf{UM(2)}}, respectively, in Figure \ref{fig:mesh}), we establish a highly refined but homogeneous mesh with a small spacing parameter $D$. Explicitly, for $s=d$, this mesh comprises $2N_1+1$ points defined as follows:
\begin{align}
    s_{i}&:= d + i D,\ \text{for }i=0, ..., N_1, \\
    s_{N_1+i}&:= d - i D,\ \text{for }i=1, ..., N_1.
\end{align}


\subsubsection{Geometric mesh:}
In order to reduce the total number of mesh points  we connect the small-$D$ linear mesh at surfaces with a mesh of exponentially increasing $s_i$. This facilitates a smooth transition from material boundaries, where fine spacing is crucial, to the center of homogeneous regions, allowing for a coarser grid. Taking \textsf{GM(3)} in Figure (\ref{fig:mesh}) as an example,  the mesh is constructed as follows: Initially, the mesh boundaries are set at $a:=d-N_1D>0$ and $b:=0$, as the scalar field exhibits the weakest slope in the middle of the vacuum region, justifying a coarser grid. The number of points within this mesh, denoted as $N_2$, is predetermined. An exponential parameter $\delta>1$ determines the positions of all other points:
\begin{align}
    s_0 &:= a \\
    s_i &:= s_{i-1} - D \delta^{i-1},\ \text{i=1, ..., $N_2$-1}.
\end{align}
In order to ensure $s_{N_2-1} = b$, $\delta$ has to be determined by solving
\begin{align}
    \sum_{i=1}^{N_2-1} D\delta^{i-1} = D \frac{1-\delta^{N_2-1}}{1-\delta}=|b-a|.
\end{align}

\subsubsection{Full mesh for the considered geometries}

\begin{figure}[!ht]
\begin{center}
\includegraphics[scale=0.86]
{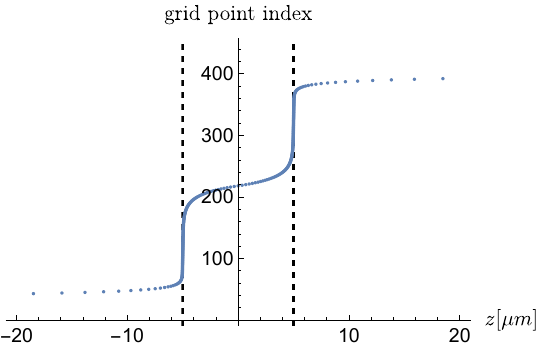}
\caption{This illustration provides an example of the grid construction realizing grid spacings between 1 nm and 19 mm with only $\sim 430$ points. The dotted lines mark the surfaces of the mirror. The mesh parameters are $D_1=D_2=1$ nm, $N_1=N_2=80$ and $N_3=30$.}
\label{fig:grid}
\end{center}
\end{figure} 

The full mesh is then formed by combining uniform and geometric meshes, illustrated in Figure \ref{fig:mesh}. Figure \ref{fig:grid} exemplifies the final result for a two mirror geometry. For one mirror and spherical geometries only the meshes GM(1), UM(1) and GM(2) need to be constructed, while for a two-mirror geometry, due to the symmetry of the setup, GM(3), UM(2) and GM(4) are chosen to be the mirror image of GM(1), UM(1) and GM(2). This results in a total of 5 parameters to define the mesh: Two spacing parameters $D_1$ and $D_2$ alongside with the parameters $N_1$ and $N_2$ two determine the number of points in the geometric meshes, as well $N_3$ to set the number of points for the uniform meshes. In most cases that we investigated, setting $D_1=D_2$, $N_1=N_2$ lead to very good numerical results.

\subsubsection{Mesh for geometries with less symmetry}

For geometries exhibiting lower symmetry, the construction of higher-dimensional grids becomes necessary. In our prior work \cite{Fischer:2023eww}, we simulated the dilaton field within a cylinder, prompting the development of a two-dimensional grid defined by the coordinates $r$ and $z$. To achieve this, we implemented the construction of two one-dimensional grids, one for the radial points $r_0,...,r_{N_1}$ representing the cross-section of the cylinder, and the other for the axial points $z_0,...z_{N_2}$ indicating the length of the cylinder. Our findings revealed that by employing two separate one-dimensional grids, constructed as elucidated earlier, we could effectively define a suitable two-dimensional grid. This grid is characterized by the points $\{(r_i,z_j)|\ 0\leq i \leq N_1,\ 0\leq j \leq N_2\}$. We anticipate that a similar construction approach will prove viable for three-dimensional geometries as well.

\subsection{Testing the algorithm for the two mirror solution of the dilaton field} \label{DilatonTests}

\begin{figure}[!ht]
\begin{center}
\includegraphics[width=1 \linewidth]
{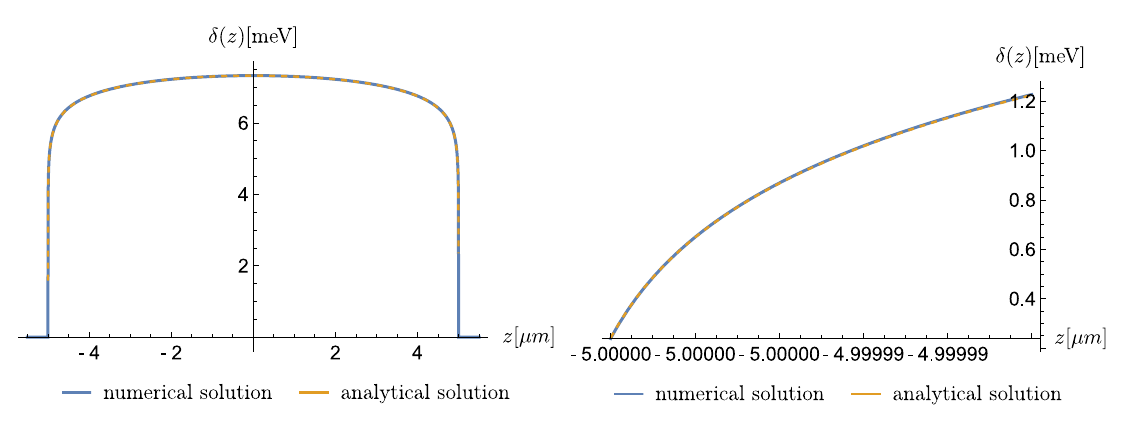}
\caption{The dilaton field defined in Eqs. (\ref{final}) and (\ref{newdef}) is plotted for $\gamma = 1$, $A_2 = 10^{45}$, $\lambda=10^{31}$, $\rho_M = 1.083\times 10^{-5}\,$MeV$^4$ and $d=\SI{5}{\micro\metre}$ and compared to the solution of the numerical algorithm. The mesh parameters are $D_1=D_2=10\,$fm, $N_1=N_2=80$ and $N_3=30$. For the given parameters $\phi_M \simeq 11.9\,$meV.}
\label{fig:Comp1}
\end{center}
\end{figure} 

\begin{figure}[!ht]
\begin{center}
\includegraphics[width=0.5 \linewidth]
{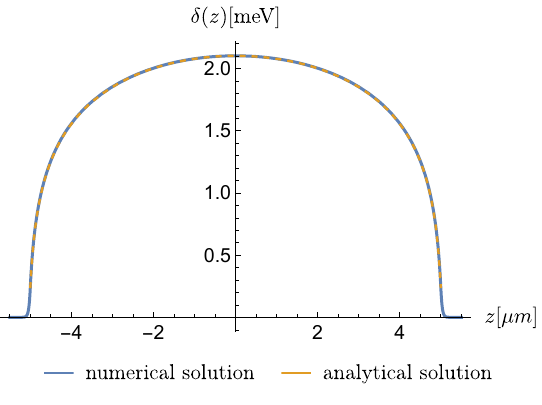}
\caption{The dilaton field defined in Eq. (\ref{final}) and Eq. (\ref{newdef}) is plotted for $\gamma=10^{24}$, $A_2 = 10^{13}$, $\lambda=10^{31}$, $\rho_M = 1.083\times 10^{-5}$  MeV$^4$ and $d=\SI{5}{\micro\metre}$ and compared to the solution of the numerical algorithm. The mesh parameters are $D_1=D_2=1$ nm, $N_1=N_2=80$ and $N_3=30$. For the given parameters $\phi_M \simeq 5.61\times 10^{23}\,$meV.}
\label{fig:Comp1Alt}
\end{center}
\end{figure} 

We demonstrate the accuracy of the proposed algorithms for the case of extreme slopes in Figure \ref{fig:Comp1} and for the case of an extremely large value of $\gamma$ in Figure \ref{fig:Comp1Alt}, by computing $\delta$ as defined in Eq. (\ref{newdef}). The differential equation (\ref{GoodEq}) for $\delta$ can be solved with the proposed algorithm by defining 

\begin{align}
    V_{\text{eff}}(\delta):= \frac{\beta(\phi_M)}{\lambda}\rho_M {\rm e}^{- \lambda \delta / m_{\text{pl}}} + \beta\left(\phi_M+\frac{\delta}{2}\right)\rho \frac{\delta}{m_{\text{pl}}}.
\end{align}

Numerical field solutions where computed with standard machine precision (16 digits), while the corresponding analytical solution for $\gamma = 10^{24}$ had to be computed with much higher precision (at least $\sim$ 30 digits) for comparison. The analytical and numerical solution closely match even inside extreme slopes. Since the analytical solution assumes $\rho_V=0$, a numerical solution is mandatory for a realistic analysis of \cannex{} and other experiments, where this restriction may not hold in general.

\subsection{Testing the algorithm for the two mirror solution of the symmetron and chameleon fields}

\begin{figure}[H]
\begin{center}
\includegraphics[width=1 \linewidth]
{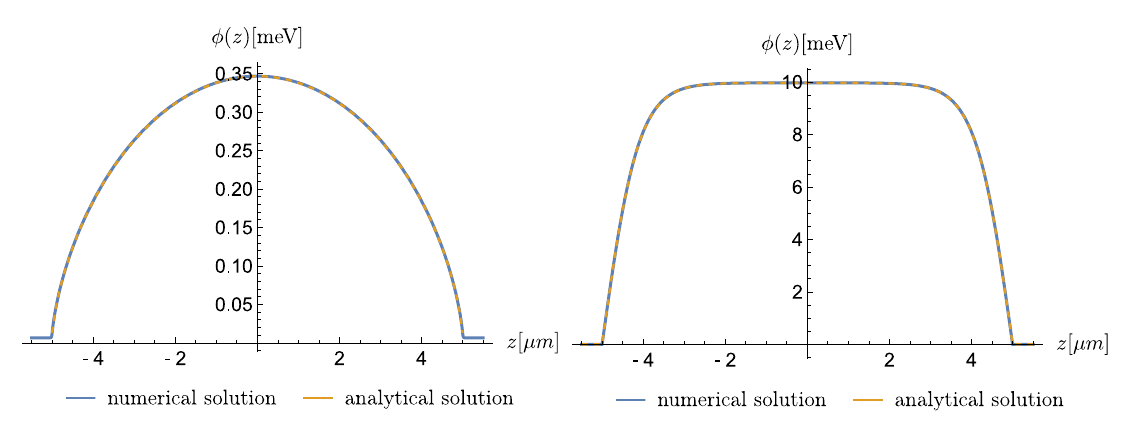}
\caption{$Left:$ The analytically exact two mirror solutions for the chameleon field for the parameters $n=1,\ \beta=4.1\times 10^{5}$ and $\Lambda = 2.4\times 10^{-9}$ is plotted alongside the solution of the numerical algorithm. $Right:$ The analytically exact two mirror solutions for the symmetron field for the parameters $\mu=10^{-6.5}\,$MeV$, M=10^{-3.2}\,$MeV and $\lambda=10^3$ is plotted alongside the solution of the numerical algorithm. The experimental parameters were set to $\rho_M = 1.083\times 10^{-5}$ MeV$^4$, $\rho_V=0$ and $d = \SI{5}{\micro\metre}$. The mesh parameters are $D_1=D_2=1\,$nm, $N_1=N_2=80$ and $N_3=30$. }
\label{fig:Comp2}
\end{center}
\end{figure} 

We also tested our code for the symmetron and chameleon model, since exact analytical solutions are available \cite{Brax:2017hna,Ivanov:2016rfs}.  Examples of the obtained numerical solutions in comparison to the respective analytical ones are shown in Figure \ref{fig:Comp2}. While the analytically exact solutions for the chameleon field is only valid for $n=1$, the provided algorithm can compute the field for arbitrary parameters.

\subsection{Testing for a one mirror and spherical geometries}

We tested that our code can reproduce the known one-mirror solution  \cite{Brax:2017hna} for the symmetron field. While analytically exact solutions of the investigated scalar fields for spheres remain elusive, inside the small $\lambda$ region, the dilaton differential equation becomes linear and can be explicitly solved \cite{Brax:2022uyh}. We checked that the numerical algorithm and the approximate analytical solution match in this scenario.

\section{Numerical solution of the stationary Schr\"odinger equation in the presence of a scalar field in one dimension} \label{6}

For quantum particles such as neutrons, scalar fields can result in an additional potential, thereby altering the energy states of the object. We discuss how the above methods can be extended to compute these energies in the one dimensional case. We assume a particle of mass $m$ inside a potential $V(z)$. The eigenenergies and eigenstates would be given by 
\begin{align}
    - \frac{1}{2 m} \frac{d^2 \Psi_n}{dz^2}(z)+V(z) \Psi_n(z)=E_n\Psi_n(z)\,.\label{SER}
\end{align}
We now introduce the effective interaction with a scalar field via the additional effective potential $V_S(z)=m\mathfrak{Q}A(\phi)$~\cite{Brax:2022uyh,Pitschmann:2020ejb} that includes the Weyl rescaling $A(\phi)$ and adds to the background potential $V^0(z)$, by setting
\begin{align}
    V(z):=V^0(z)+V_S(z).
\end{align}
The factor $\mathfrak{Q}$ is a `screening charge' that takes into account how strongly the particle screens. For unscreened particles it takes the value 1, for fully screened particles 0. These factors are approximately known for spherical geometries ~\cite{Brax:2017hna,Brax:2022uyh,Fischer:2023eww}.

\subsection{Discretization}

The first step in solving this problem is to compute $\phi$ and hence $V_S$ as described in the previous section. Secondly, the eigenvalue problem \ref{SER} can be discretized in an analogous method as the scalar field equations of motion using Eq. (\ref{dx}). This results in the discrete Eigenvalue problem 

\begin{align}
    H_D \Psi= E \Psi,
\end{align}
with $\Psi=(\Psi_1,...,\Psi_N)$ and

\begin{align}
H_{D,ij} = \begin{cases}
      -\displaystyle\frac{1}{2m} \frac{2}{h_i(h_i+h_{i-1})} &\text{, if } j = i+1\\
     \displaystyle-\frac{1}{2m} \frac{2}{h_{i-1}(h_i+h_{i-1})} &\text{, if } j = i-1\\
     -H_{ii} - H_{i,i-1} + V_i&\text{, if } j = i \\
     0 &\text{, else}\>.
    \end{cases} 
\end{align}

Boundary conditions for $\Psi$ can be implemented by requiring $\Psi_0=\Psi_{N+1}=0$, in analogy to the scalar field cases. For non-uniform meshes the discrete hamiltonian is in general not symmetric, and real eigenvalues of the discrete problem might not even exist. Symmetry can be restored by applying a suitable transformation as detailed in \cite{tan1990self}. The mesh construction will in general depend on both $V^0$ and  $V_S$. To solve the discrete eigenvalue problem we used Mathematica's Eigensystem function that numerically computes all eigenvalues and eigenfunctions obtainable with the given resolution of the mesh.

\subsection{Testing the algorithm for the one mirror geometry}

We now test our algorithm for solving the Schr\"odinger equation described above at a test case. For this, we assume an ultra cold neutron bouncing inside a vacuum region above a horizontal mirror located at $z<0$ and trapped in the gravitational potential $V^0=mgz$ of the Earth. This calculation is relevant to the \qbounce{} experiment, as described later. We assume that the neutron is a (possibly screened) spherical particle with radius 0.5\,fm in order to take its screening into account. 

\subsubsection{Mesh construction}

Numerical computation of energies and wave functions necessitates the construction of two distinct meshes:

Firstly, $V_S$ must be computed by solving the equations of motion for the scalar field (considering a one-mirror geometry in this case). The mesh construction process is detailed in Section \ref{MeshCon}.

Secondly, a suitable mesh needs to be generated to solve the stationary Schrödinger equation in the presence of the computed scalar field. In our investigations, we have observed that in many cases, it is inadequate to adopt the mesh constructed for dilaton solutions as described in Section \ref{MeshCon}. This is because the neutron wave functions tend to exhibit a markedly different shape compared to the dilaton field solutions. Specifically, the wave functions may either compress themselves into the initial strong slope of the dilaton field -- a behavior that can be adequately captured by a single uniform mesh of very small extent -- or they may exhibit minimal influence from the dilaton field, relaxing to the Newtonian case. In the latter scenario, a uniform mesh with a larger extent proves effective.

For hybrid cases, we have found that utilizing two uniform meshes is appropriate: one designed to resolve the initial slope of the dilaton field, and a second one tailored to capture the shape of the wave functions, following the Newtonian contribution of the potential.

\subsubsection{Tests}

It is noteworthy that the Newtonian case can be solved analytically, as detailed in \cite{MarioHabil}. The solution is expressed as:
\begin{align}
 \Psi_n(z) &= \frac{\text{Ai}(\frac{z-z_n}{z_0})}{\sqrt{z_0}\text{Ai}'(-\frac{z_n}{z_0})}, \label{wavefunc}\\
 E_n &= \sqrt[3]{\frac{m g^2}{2}}\times \text{Ai}(0,n) \label{energies}.
 \end{align}

Here, $\Psi_n$ represents the $n$-th energy state with energy $E_n$, $z_0=  \sqrt[3]{\frac{1}{2m^2g}}=\SI{5.9}{\micro\metre}$ determines the extent of the wave functions, $ z_n = \frac{E_n}{m g}$, $\text{Ai}$ denotes the Airy functions (see e.g., \cite{olivier2010airy}), and $\text{Ai}(0,n)$ refers to the $n$-th root of the Airy function.

We conduct tests on our algorithm for three different cases, depending on the extent to which the dilaton field perturbs the wave functions and energies. Heuristically, we define the strength of the perturbation as
\begin{align*}
    \text{pert}(n):= \left|\frac{E_n-E_n^0}{E_n^0}\right|.
\end{align*}

\begin{itemize}
    \item For a negligible perturbation ($\text{pert}(n) \ll 1$), we expect the wave functions and engeries to converge to the Newtonian case, which is known exactly.
    \item For a strong perturbation ($\text{pert}(n) \gg 1$), $E_n$ is in general not known analytically. However, for parameters where the dilaton field is long-ranged it can be taylor expanded over the extent of the wave function we can use 
    \begin{align}
    V_S(z) \simeq V_S(0)+ \frac{dV_S}{dz}(0)z.
\end{align}
Plugging this expression into Eq. (\ref{SER}) leads again to the unperturbed Newtonian case with $g$ replaced by $g_{\text{eff}}=g+\frac{1}{m}\frac{dV_S}{dz}(0)$. The energies and wave functions can hence be computed analytically from Eqs. (\ref{wavefunc}) and (\ref{energies}). 
\end{itemize}
\begin{itemize}
    \item For a mild perturbation ($\text{pert}(n) \simeq 0.1$), $E_n$ can approximately be computed with perturbation theory using (see e.g. \cite{landau1991quantenmechanik})
\begin{align}
    E_n = E^{(0)}_n+\int_{-\infty}^{\infty}dz |\Psi^{(0)}_n|^2 V_S(z).
\end{align} 
\end{itemize}

\begin{figure}[H]
\begin{center}
\includegraphics[width=1 \linewidth]
{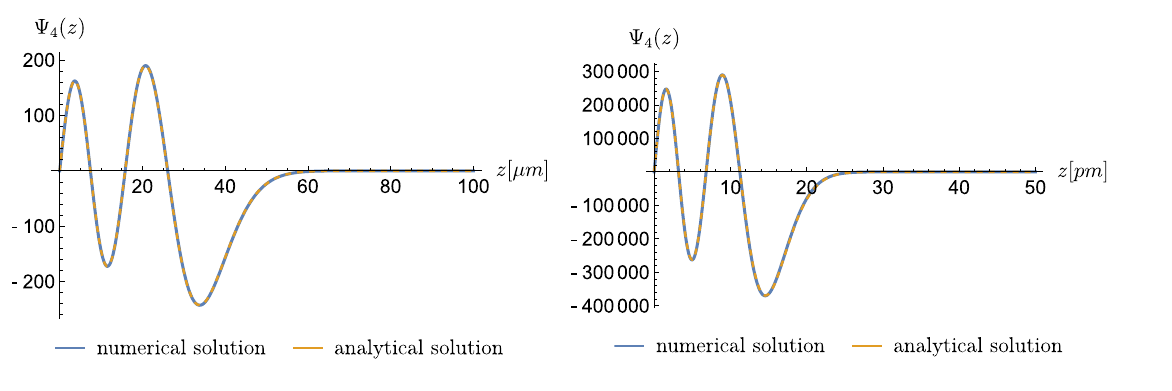}
\caption{The analytical solution is plotted alongside the numerical solution for the fourth energy state, for a negligible dilaton perturbation (left) and a strong dilaton perturbation (right). The parameters are detailed in Table \ref{Table2}. A single uniform mesh was used in each case, consisting of 400 points each and a cutoff at \SI{100}{\micro\metre} (left), and at 50\,pm (right).}
\label{fig:CompANA3}
\end{center}
\end{figure}

A comparison between numerical and exact solutions for the first two cases is shown in Figure \ref{fig:CompANA3}, and a comparison of the energies for all cases is presented in Table \ref{Table2}. \\

We note that the strongly perturbed case shown in Figure \ref{fig:CompANA3} (right) corresponds to a strong long-ranged dilaton field that results in $g_{\text{eff}}=1.23\times 10^{20}\frac{\text{m}}{\text{s}^2}$ and hence $\sim z_0 = \sqrt[3]{{1}/(2m^2g_{\text{eff}})}=2.53\,$pm. The wave function is strongly compressed. While this case may not be physically interesting (given that the observed extent of the wave functions is much larger in the \qbounce{} experiment), it naturally arises when computing parameter constraints for the dilaton field. This is evident when comparing the parameters in Table \ref{Table2} with the constraints in Figure \ref{fig:PS}. Parameters that predict an experimental outcome that has not been observed can be constrained, requiring an accurate calculation even for such cases. We include this scenario as it is one of the few possible cases that we can solve exactly and use to test the numerical method.


\begin{table}[ht]
\centering
\renewcommand{\arraystretch}{1.7}
\begin{tabular}{cccccc}
\hline\hline
Perturbation & $V_0[\text{MeV}^4]$ & $A_2$ & $\lambda$ & $\left|\frac{E_{1,\text{ana}}-E_{1,\text{num}}}{E_{1,\text{ana}}}\right|$ & $\left|\frac{E_{4,\text{ana}}-E_{4,\text{num}}}{E_{4,\text{ana}}}\right|$\\
\hline
Negligible & 10 & $10^{40}$ & $10^{40}$ & $7.1 \times 10^{-5}$ & $2.1 \times 10^{-4}$\\
Weak  & 10 & $10^{40}$ & $10^{31}$ & $4.5 \times 10^{-5}$ & $1.3 \times 10^{-4}$\\
Strong & 10 & $10^{40}$ & $10^{23}$ & $5.9 \times 10^{-4}$ & $2.9 \times 10^{-3}$\\
\hline\hline
\end{tabular}
\caption{Comparison of eigenenergies computed using the analytically exact solution and perturbation theory (subscript `ana') and those computed using the numerical algorithm (subscript `num').\label{Table2} 
}
\end{table}

\section{Application to experimental constraints}\label{7}

In this section we discuss how the provided methods can be applied to derive constraints for different experiments.

\subsection{Computing energy shifts induced by scalar fields in the \qbounce{} experiment}

\begin{figure}[H]
\begin{center}
\includegraphics[width=0.5 \linewidth]
{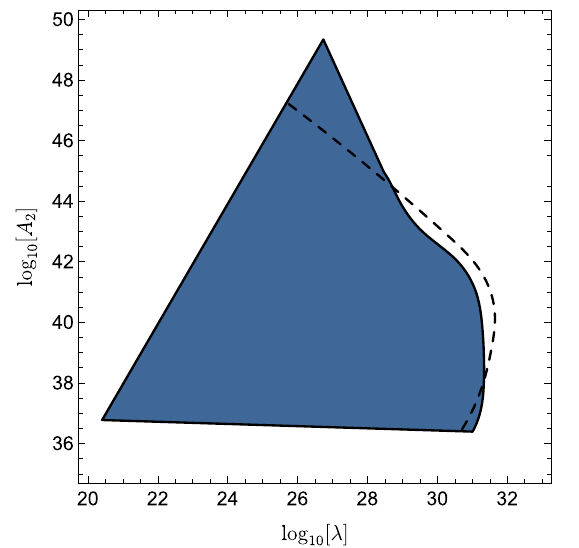}
\caption{The blue area shows the \qbounce{} constraints for the dilaton field for $\gamma=1$ obtained from the numerical methods provided in this paper (assuming the neutron is a sphere with radius 0.5 fm). The dashed line indicates the constraints computed from perturbation theory alone and an analytical approximation of the dilaton field, presented in \cite{Brax:2022uyh}.}
\label{fig:CompANA4}
\end{center}
\end{figure}

\qbounce{} measures energy transitions of ultracold neutrons in the gravitational field of the Earth. In the Rabi implementation of \qbounce{}~\cite{Cronenberg:2018qxf}, the neutrons pass three regions. The first region consists of a state selector that scatters higher energy states out of the experiment. Only neutrons in the lowest few energy states $E_1=1.41\,$peV, $E_2=2.46\,$peV, remain. In the second region, neutrons can be excited to a higher energy state (for example $E_3=3.32$ peV or $E_4=4.08$ peV) by a vibrating neutron mirror with tunable frequency $\omega$. In the last region higher energy states are again scattered out of the system. The experiment can resolve energy differences with a resolution of $2\times 10^{-15}$ eV. Scalar fields would result in an additional potential that can (to good approximation) be obtained from the one mirror solution. Constraints can hence be placed on scalar field parameters for which
\begin{align}
    |(E^s_1-E^s_n)-(E^{0}_1-E^{0}_n)|\geq 2 \times 10^{-15}\text{eV},
\end{align}
for $n\in \{3,4\}$, where superscript 0 refers to the unperturbed Newtonian case and $s$ to energies in the presence of a scalar field. To demonstrate the large improvement over previous methods only relying on an analytical approximation of the dilaton field and perturbation theory \cite{Brax:2022uyh}, we demonstrate in Figure \ref{fig:CompANA4} that the computed parameter constraints change in some regions by several orders of magnitude due to the more accurate methods provided in this paper. From these results it is clear that the previous analysis of \qbounce{} in the context of the symmetron field that solely relied on perturbation theory \cite{Jenke:2020obe} could be considerably improved with the method presented herein.

\subsection{Computing observables for Lunar Laser Ranging}

Lunar Laser Ranging (LLR) serves as a sophisticated method for the precise determination of the Earth-Moon distance, that can be used to test general relativity. One such constraint pertains to potential violations of the equivalence principle in the gravitational accelerations $a$ of the Earth (\earth) and Moon (\leftmoon) towards the Sun (\sun) \cite{Brax:2022uyh,nordtvedt2001lunar}. This constraint is expressed as follows:
\begin{align}
\delta_{\text{em}}\simeq \frac{|\vec{a}_{\phi \earth}-\vec{a}_{\phi \leftmoon}|}{|\vec{a}_G|}  \leq 2 \times 10^{-13}\>. \label{C1}
\end{align}

Here, $\vec{a}_\phi$ denotes the acceleration of the celestial body in the index towards the Sun caused by a screened scalar field, in addition to the Newtonian acceleration $\vec{a}_G$. Another significant constraint arises from deviations in the inverse square law, leading to a precession of the lunar perigee. This constraint is given by:
\begin{align}
\left|\frac{\delta \Omega}{\Omega}\right| &\simeq \left| \frac{R^2}{G M_{\earth}}\,\big(\delta f(R) + \frac{R}{2}\, \delta f'(R)\big)\right|\nonumber\\
& \leq 6.23833 \times 10^{-12}\>. \label{C2}
\end{align}
In this inequality $R$ represents the maximum Earth-Moon separation (apogee), $M_{\earth}$ is the mass of the Earth, and $\delta f(R)$ signifies the centripetal acceleration caused by a scalar field dependent on the separation $R$. 
A scalar field $\phi$ would lead to (see Eq. (\ref{force})) 
\begin{align}
    \delta_{\text{em}} &\simeq  \frac{1}{|\vec{a}_G|}\left|(\mathfrak{Q}_{\earth}-\mathfrak{Q}_{\leftmoon})\beta\big(\phi(r)\big) \frac{\nabla \phi (r)}{m_{\text{pl}}}\right|_{r=1\,\text{AU}}\>,\\
    \delta f(r) &\simeq  \mathfrak{Q}_{\leftmoon}\beta\big(\phi(r)\big) \frac{\nabla \phi (r)}{m_{\text{pl}}},
\end{align}
where we again multiplied the forces of the test particles with the screening charge of the Earth / Moon. 

\subsection{Computing the pressure in the \cannex{} experiment}

In the \cannex{} experiment, two parallel plates are located at $z<-d$ and $d<z<d+D$ with density $\rho_M$ with vacuum in between them. The experiment is designed to measure the pressure due to the Casimir effect as well as hypothetical fifth forces between truly parallel plates. Scalar fields would induce such an additional pressure along the $z$-axis, as was computed in Ref.~\cite{Fischer:2023koa},

\begin{align}
P_z = \frac{\rho_M}{\rho_M-\rho_V}\,\big(V_{\text{eff}}(\phi_V,\rho_V)-V_{\text{eff}}(\phi_0,\rho_V)\big)\>, \label{newpress}
\end{align}
where $\phi_0$ is the value of the scalar field just in between both plates, as elaborated on in Section \ref{3}. This value can be computed from the two mirror solution. The sensitivity of this experiment for the dilaton pressure is predicted to be around 1\,nPa, hence constraining any scalar field parameters that would lead to $|P_z| \geq 1$ nPa. We note that a rigorous limit calculation would involve a statistical test (e.g. a chisquare or regression analysis) quantifying the probability for the dilaton to exist or not  for all parameter combinations. However, given the vast uncertainties and approximations in the present theory~\cite{Fischer:2023koa} we only aim at `order-of-magnitude precision' limits, for which a simple comparison with the experimental sensitivity is sufficient.

\section{Conclusion}\label{8}

In this study, we have derived the first exact field solution for the environment-dependent dilaton theory. Our findings demonstrate that a strategic reparameterization of the model, coupled with a transformation of the equations of motion, effectively mitigates the numerical challenges induced by the exponential self-coupling of the field, even when dealing with demanding parameter sets. Additionally, we have showcased the efficacy of a non-uniform finite difference discretization in resolving the distinct characteristics of extreme slopes inherent to this model -- whether in one- and two-mirror configurations, or in spherical geometries. 

Furthermore, we introduced a complementary approach to solve the 1D Schrödinger equation in the presence of a scalar field, demonstrating a significant improvement over the application of perturbation theory. Our results suggest that the limits on screened scalar fields computed for many other experiments may need to be re-evaluated carefully, taking into consideration the approximations and limits of applicability of the used theoretical approach. The versatility of our numerical algorithm extends beyond the dilaton field theory, as it can be successfully applied to other models such as symmetrons and chameleons. In a future work, we will use the methods herein to generalize existing constraints to other models.

We provide a Mathematica implementation of all methods presented in this paper at \cite{H.Fischer}.

\section{Acknowledgments} \label{9}
We thank Mario Pitschmann for fruitful discussions. This research was funded in whole or in part by the Austrian Science Fund (FWF) P 34240-N. For open access purposes, the author has applied a CC BY public copyright license to any author accepted manuscript version arising from this submission. During the preparation of this work, the authors used ChatGPT 3.5 to enhance the language quality. After utilizing this tool, the authors reviewed and edited the content as needed, taking full responsibility for the publication's content.

\appendix
\nopagebreak

\section{Newton's method for the finite difference scheme}\label{AppSM1}

We show how to solve the discretized differential equation (Eq. \ref{DiscreteEq}) and explain in detail how to apply boundary conditions.
Defining 
\begin{align}
F: \mathbb{R}^N &\rightarrow  \mathbb{R}^N \notag\\
F_i(\phi_1, ..., \phi_N) &=\frac{2(\phi_{i+1}-\phi_{i})}{h_i(h_i+h_{i-1})}-\frac{2(\phi_{i}-\phi_{i-1})}{h_{i-1}(h_i+h_{i-1})}+ a\frac{2}{s_i}\big[\frac{h_i(\phi_i-\phi_{i-1})}{h_{i-1}(h_i+h_{i-1})}+\frac{h_{i-1}(\phi_{i+1}-\phi_{i})}{h_{i}(h_i+h_{i-1})}\big] - V_{\text{eff}, \phi}(\phi_i, \rho_i), \label{Newton}
\end{align}
results in an $N$ dimensional root finding problem for $F$. The values $\phi_0$ and $\phi_{N+1}$ have to be fixed to set the boundary conditions. The equation 
\begin{align}
    F(\phi_1, ..., \phi_N) = (0, ...,0)^T
\end{align}
can be solved using Newton's method (see e.g. \cite{press2007numerical}).
Starting from an initial guess $\phi^0 = (\phi_1^0, ..., \phi_N^0)$ we linearize $F$ around that guess and solve

\begin{align}
F(\phi^0) + J_{F}(\phi^0) (\phi^1-\phi^0) = 0, \label{convergence}
\end{align}

with the Jacobi Matrix $(J_F)_{ij}:= \frac{\partial F_i}{\partial \phi_j}.$ This defines the improved guess $\phi^1$. This procedure is iterated. In the n'th iteration the linear system 

\begin{align}
A^{n-1}\phi^n &= b^{n-1}, \notag\\
A^{n-1}:&= J_{F}(\phi^{n-1}), \notag\ \\
 b^{n-1}:&= - F(\phi^{n-1}) +  J_{F}(\phi^{n-1}) \phi^{n-1}. \label{NEWTON}
\end{align}

has to be solved. For $i=2, ..., N-1$, $A^{n-1}$ and $b^{n-1}$ are given by

\begin{align}
A^{n-1}_{ij} = \begin{cases}
       \frac{2}{h_i(h_i+h_{i-1})} + a\frac{2}{s_i}\frac{h_{i-1}}{h_i(h_i+h_{i-1})} &\text{, if } j = i+1\\
       \frac{2}{h_{i-1}(h_i+h_{i-1})}-a\frac{2}{s_i}\frac{h_i}{h_{i-1}(h_i+h_{i-1})} &\text{, if } j = i-1,\\
     -\frac{2}{h_i(h_i+h_{i-1})} - \frac{2}{h_{i-1}(h_i+h_{i-1})} +a\frac{2}{s_i}\big[\frac{h_i}{h_{i-1}(h_i+h_{i-1})}-\frac{h_{i-1}}{h_{i}(h_i+h_{i-1})}\big]
     -V_{\text{eff},\phi \phi}(\phi^{n-1}_i, \rho_i) &\text{, if } j = i, \\
     0 &\text{, else } ,
    \end{cases} 
\end{align}
and

\begin{align}
b^{n-1}_{i} =V_{\text{eff}, \phi}(\phi^{n-1}_i; \rho_i) -V_{\text{eff}, \phi \phi}(\phi^{n-1}_i; \rho_i)\phi^{n-1}_i.
\end{align}

The remaining values have to take the boundary conditions into account.

\subsubsection{Boundary conditions for the one and two mirror case}

As boundary conditions we demand that the fields minimize their effective potential at the boundary, hence $\phi_0=\phi_M$, and $\phi_{N+1} \in \{\phi_V, \phi_M\}$ for the one and two mirror case respectively. This results in

\begin{align}
A^{n-1}_{1,1}&= - \frac{2}{h_1^2} -V_{\text{eff},\phi \phi}(\phi^{n-1}_1; \rho_1), \\
A^{n-1}_{1,2}&= \frac{1}{h_1^2}, \\
A^{n-1}_{N,N-1}&= \frac{1}{h_{N-1}^2},\\
A^{n-1}_{N,N}&= - \frac{2}{h_{N-1}^2}  -V_{\text{eff},\phi \phi}(\phi^{n-1}_N, \rho_N), \\
b^{n-1}_1 &= V_{\text{eff}, \phi}(\phi^{n-1}_1; \rho_1) -V_{\text{eff}, \phi \phi}(\phi^{n-1}_1; \rho_1)\phi^{n-1}_1- \frac{\phi_0}{h_1^2}, \\
b^{n-1}_N&= V_{\text{eff}, \phi}(\phi^{n-1}_N; \rho_N) -V_{\text{eff}, \phi \phi}(\phi^{n-1}_N; \rho_N)\phi^{n-1}_N - \frac{\phi_{N+1}}{h_{N-1}^2}.
\end{align}

We set $h_0:=h_1$, $h_N:=h_{N-1}$. 

\subsubsection{Boundary conditions for a sphere}

\textbf{r=0:}\\
For a spherical geometry $\frac{d\phi}{dr}|_{r=0}\>=0$ has to hold to handle the coordinate singularity at $r=0$ in Eq. (\ref{DEQ2}).

We define a ghost point at $r_0:=-r_1<0$ with no physical meaning, and define
\begin{align}
    \phi_0:= \phi_1. \label{symmetry}
\end{align}

 While $\phi_{1/2}:=\phi(0)$ is not part of the discretization to avoid the singularity, the vanishing of $\frac{d\phi}{dr}|_{r=0}\>$ is guaranteed by Eq. (\ref{approx}), if the neighboring points of  $\phi_{1/2}$ are chosen to be $\phi_{0}$ and $\phi_{1}$.  This idea to avoid the singularity has essentially been put forward in \cite{mohseni2000numerical}.
 Using (\ref{symmetry}) for $i=1$ and $r_1=h_0/2$ results in 

 \begin{align}
     A_{1,1}&:= - \frac{6}{h_1(h_1+h_0)}-V_{\text{eff},\phi \phi}(\phi^{n-1}_1, \rho_1),\\
      A_{1,2}&:= \frac{6}{h_1(h_1+h_0)},\\
      b_1&:=V_{\text{eff}, \phi}(\phi^{n-1}_1; \rho_1) -V_{\text{eff}, \phi \phi}(\phi^{n-1}_1; \rho_1)\phi^{n-1}_1.
 \end{align}

\textbf{$r=\text{cutoff}$}:\\

By setting $h_N:=h_{N-1}$ and using $\phi_{N+1}=\phi_V$ we get 

\begin{align}
   A^{n-1}_{N,N-1}&= \frac{1}{h_{N-1}^2}-\frac{1}{r_N h_{N-1}},\\
A^{n-1}_{N,N}&= - \frac{2}{h_{N-1}^2}  -V_{\text{eff},\phi \phi}(\phi^{n-1}_N, \rho_N), \\ 
b^{n-1}_N&= V_{\text{eff}, \phi}(\phi^{n-1}_N; \rho_N) -V_{\text{eff}, \phi \phi}(\phi^{n-1}_N; \rho_N)\phi^{n-1}_N - \frac{\phi_{V}}{h_{N-1}^2}- \frac{\phi_{V}}{r_N h_{N-1}}.
\end{align}

\subsubsection{Convergence}

To monitor convergence we use the euclidean norm

\begin{align}
    ||\phi^n||_2 := \sum_{i=0}^{N+1}(\phi^n_i)^2,
\end{align}

and stop iterations when

\begin{align}
    \frac{||\phi^n-\phi^{n-1}||_2}{||\phi^{n-1}||} < \varepsilon,
\end{align}
for a sufficiently small value of $\varepsilon$. In the implementation we used Mathematica for solving the linear systems of equation. The remaining task is to find an initial guess.

\subsubsection{Initial guess}
Ensuring the convergence of Newton’s method hinges on the judicious selection of an initial guess, denoted as $\phi^0$. In our exploration of the models and geometries within this study, we have identified three reliable seeds that consistently lead to convergence.

For the dilaton and chameleon models, either of the following trivial guesses guarantees convergence:

\begin{itemize}
    \item $\phi^0_i:=\phi_M$\\
    \item $\phi^0_i:=\phi_{\rho(s_i)}$
\end{itemize}

In the symmetron model, we initiate the process by expanding the effective potential around its potential minimum, resulting in a linear differential equation:

\begin{align}
    \frac{d^2\phi}{ds^2}+a \frac{2}{s}\frac{d\phi}{ds}=  V_{\text{eff}, \phi \phi}(\phi_{\rho(s)})\left(\phi(s)-\phi_{\rho(s)}\right). \label{approx3}
\end{align}
The solution to this equation serves as a reliable initial guess for the subsequent full nonlinear differential equation governing the symmetron field. Eq. (\ref{approx3})  is solved using the same algorithm described earlier, defining $V_{\text{eff,approx}}:= \frac{1}{2}V_{\text{eff}, \phi \phi}(\phi_{\rho(s)})\left(\phi(s)-\phi_{\rho(s)}\right)^2,$ and solving Eq. (\ref{DEQ2}) for $V_{\text{eff,approx}}$.  Importantly, due to the linearity of Eq. (\ref{approx3}) Newton’s method, while not strictly required after discretization, can still be formally employed. An arbitrary initial guess can be specified, providing a convenient means to avoid additional coding for solving linear differential equations. As linearizing an already linear equation has no effect, Newton’s method halts after a single iteration.

\bibliography{biblio.bib}

\providecommand{\href}[2]{#2}\begingroup\raggedright\begin{thebibliography}{10}

\bibitem{SupernovaCosmologyProject:1997zqe}
{\scshape Supernova Cosmology Project} collaboration, \emph{{Discovery of a
  supernova explosion at half the age of the Universe and its cosmological
  implications}}, \href{https://doi.org/10.1038/34124}{\emph{Nature} {\bfseries
  391} (1998) 51} [\href{https://arxiv.org/abs/astro-ph/9712212}{{\ttfamily
  astro-ph/9712212}}].

\bibitem{SupernovaSearchTeam:1998fmf}
{\scshape Supernova Search Team} collaboration, \emph{{Observational evidence
  from supernovae for an accelerating universe and a cosmological constant}},
  \href{https://doi.org/10.1086/300499}{\emph{Astron. J.} {\bfseries 116}
  (1998) 1009} [\href{https://arxiv.org/abs/astro-ph/9805201}{{\ttfamily
  astro-ph/9805201}}].

\bibitem{SupernovaSearchTeam:1998bnz}
{\scshape Supernova Search Team} collaboration, \emph{{The High Z supernova
  search: Measuring cosmic deceleration and global curvature of the universe
  using type Ia supernovae}},
  \href{https://doi.org/10.1086/306308}{\emph{Astrophys. J.} {\bfseries 507}
  (1998) 46} [\href{https://arxiv.org/abs/astro-ph/9805200}{{\ttfamily
  astro-ph/9805200}}].

\bibitem{Frieman:2008sn}
J.~Frieman, M.~Turner and D.~Huterer, \emph{{Dark Energy and the Accelerating
  Universe}},
  \href{https://doi.org/10.1146/annurev.astro.46.060407.145243}{\emph{Ann. Rev.
  Astron. Astrophys.} {\bfseries 46} (2008) 385}
  [\href{https://arxiv.org/abs/0803.0982}{{\ttfamily 0803.0982}}].

\bibitem{Sola:2013gha}
J.~Sola, \emph{{Cosmological constant and vacuum energy: old and new ideas}},
  \href{https://doi.org/10.1088/1742-6596/453/1/012015}{\emph{J. Phys. Conf.
  Ser.} {\bfseries 453} (2013) 012015}
  [\href{https://arxiv.org/abs/1306.1527}{{\ttfamily 1306.1527}}].

\bibitem{Joyce:2014kja}
A.~Joyce, B.~Jain, J.~Khoury and M.~Trodden, \emph{{Beyond the Cosmological
  Standard Model}},
  \href{https://doi.org/10.1016/j.physrep.2014.12.002}{\emph{Phys. Rept.}
  {\bfseries 568} (2015) 1} [\href{https://arxiv.org/abs/1407.0059}{{\ttfamily
  1407.0059}}].

\bibitem{Khoury:2003rn}
J.~Khoury and A.~Weltman, \emph{{Chameleon cosmology}},
  \href{https://doi.org/10.1103/PhysRevD.69.044026}{\emph{Phys. Rev. D}
  {\bfseries 69} (2004) 044026}
  [\href{https://arxiv.org/abs/astro-ph/0309411}{{\ttfamily
  astro-ph/0309411}}].

\bibitem{Khoury:2013tda}
J.~Khoury, \emph{{Les Houches Lectures on Physics Beyond the Standard Model of
  Cosmology}},  \href{https://arxiv.org/abs/1312.2006}{{\ttfamily 1312.2006}}.

\bibitem{Brax:2012jr}
P.~Brax, C.~Burrage and A.-C.~Davis, \emph{{Screening fifth forces in k-essence
  and DBI models}},
  \href{https://doi.org/10.1088/1475-7516/2013/01/020}{\emph{JCAP} {\bfseries
  01} (2013) 020} [\href{https://arxiv.org/abs/1209.1293}{{\ttfamily
  1209.1293}}].

\bibitem{Brax:2014wla}
P.~Brax and P.~Valageas, \emph{{K-mouflage Cosmology: the Background
  Evolution}}, \href{https://doi.org/10.1103/PhysRevD.90.023507}{\emph{Phys.
  Rev. D} {\bfseries 90} (2014) 023507}
  [\href{https://arxiv.org/abs/1403.5420}{{\ttfamily 1403.5420}}].

\bibitem{Damour:1994zq}
T.~Damour and A.M.~Polyakov, \emph{{The String dilaton and a least coupling
  principle}}, \href{https://doi.org/10.1016/0550-3213(94)90143-0}{\emph{Nucl.
  Phys. B} {\bfseries 423} (1994) 532}
  [\href{https://arxiv.org/abs/hep-th/9401069}{{\ttfamily hep-th/9401069}}].

\bibitem{Burrage:2017qrf}
C.~Burrage and J.~Sakstein, \emph{{Tests of Chameleon Gravity}},
  \href{https://doi.org/10.1007/s41114-018-0011-x}{\emph{Living Rev. Rel.}
  {\bfseries 21} (2018) 1} [\href{https://arxiv.org/abs/1709.09071}{{\ttfamily
  1709.09071}}].

\bibitem{Hinterbichler:2010es}
K.~Hinterbichler and J.~Khoury, \emph{{Symmetron Fields: Screening Long-Range
  Forces Through Local Symmetry Restoration}},
  \href{https://doi.org/10.1103/PhysRevLett.104.231301}{\emph{Phys. Rev. Lett.}
  {\bfseries 104} (2010) 231301}
  [\href{https://arxiv.org/abs/1001.4525}{{\ttfamily 1001.4525}}].

\bibitem{Brax:2010gi}
P.~Brax, C.~van~de Bruck, A.-C.~Davis and D.~Shaw, \emph{{The Dilaton and
  Modified Gravity}},
  \href{https://doi.org/10.1103/PhysRevD.82.063519}{\emph{Phys. Rev. D}
  {\bfseries 82} (2010) 063519}
  [\href{https://arxiv.org/abs/1005.3735}{{\ttfamily 1005.3735}}].

\bibitem{Brax:2018iyo}
P.~Brax, C.~Burrage and A.-C.~Davis, \emph{{Laboratory constraints}},
  \href{https://doi.org/10.1142/S0218271818480097}{\emph{Int. J. Mod. Phys. D}
  {\bfseries 27} (2018) 1848009}.

\bibitem{Burrage:2014oza}
C.~Burrage, E.J.~Copeland and E.A.~Hinds, \emph{{Probing Dark Energy with Atom
  Interferometry}},
  \href{https://doi.org/10.1088/1475-7516/2015/03/042}{\emph{JCAP} {\bfseries
  03} (2015) 042} [\href{https://arxiv.org/abs/1408.1409}{{\ttfamily
  1408.1409}}].

\bibitem{Hamilton:2015zga}
P.~Hamilton, M.~Jaffe, P.~Haslinger, Q.~Simmons, H.~M\"uller and J.~Khoury,
  \emph{{Atom-interferometry constraints on dark energy}},
  \href{https://doi.org/10.1126/science.aaa8883}{\emph{Science} {\bfseries 349}
  (2015) 849} [\href{https://arxiv.org/abs/1502.03888}{{\ttfamily
  1502.03888}}].

\bibitem{Burrage:2015lya}
C.~Burrage and E.J.~Copeland, \emph{{Using Atom Interferometry to Detect Dark
  Energy}}, \href{https://doi.org/10.1080/00107514.2015.1060058}{\emph{Contemp.
  Phys.} {\bfseries 57} (2016) 164}
  [\href{https://arxiv.org/abs/1507.07493}{{\ttfamily 1507.07493}}].

\bibitem{Upadhye:2012rc}
A.~Upadhye, \emph{{Symmetron dark energy in laboratory experiments}},
  \href{https://doi.org/10.1103/PhysRevLett.110.031301}{\emph{Phys. Rev. Lett.}
  {\bfseries 110} (2013) 031301}
  [\href{https://arxiv.org/abs/1210.7804}{{\ttfamily 1210.7804}}].

\bibitem{Brax:2017hna}
P.~Brax and M.~Pitschmann, \emph{{Exact solutions to nonlinear symmetron
  theory: One- and two-mirror systems}},
  \href{https://doi.org/10.1103/PhysRevD.97.064015}{\emph{Phys. Rev. D}
  {\bfseries 97} (2018) 064015}
  [\href{https://arxiv.org/abs/1712.09852}{{\ttfamily 1712.09852}}].

\bibitem{Pitschmann:2020ejb}
M.~Pitschmann, \emph{{Exact solutions to nonlinear symmetron theory: One- and
  two-mirror systems. II.}},
  \href{https://doi.org/10.1103/PhysRevD.103.084013}{\emph{Phys. Rev. D}
  {\bfseries 103} (2021) 084013}
  [\href{https://arxiv.org/abs/2012.12752}{{\ttfamily 2012.12752}}].

\bibitem{Cronenberg:2018qxf}
G.~Cronenberg, P.~Brax, H.~Filter, P.~Geltenbort, T.~Jenke, G.~Pignol et~al.,
  \emph{{Acoustic Rabi oscillations between gravitational quantum states and
  impact on symmetron dark energy}},
  \href{https://doi.org/10.1038/s41567-018-0205-x}{\emph{Nature Phys.}
  {\bfseries 14} (2018) 1022}
  [\href{https://arxiv.org/abs/1902.08775}{{\ttfamily 1902.08775}}].

\bibitem{Jenke:2020obe}
T.~Jenke, J.~Bosina, J.~Micko, M.~Pitschmann, R.~Sedmik and H.~Abele,
  \emph{{Gravity resonance spectroscopy and dark energy symmetron fields:
  qBOUNCE experiments performed with Rabi and Ramsey spectroscopy}},
  \href{https://doi.org/10.1140/epjs/s11734-021-00088-y}{\emph{Eur. Phys. J.
  ST} {\bfseries 230} (2021) 1131}
  [\href{https://arxiv.org/abs/2012.07472}{{\ttfamily 2012.07472}}].

\bibitem{Lemmel:2015kwa}
H.~Lemmel, P.~Brax, A.N.~Ivanov, T.~Jenke, G.~Pignol, M.~Pitschmann et~al.,
  \emph{{Neutron Interferometry constrains dark energy chameleon fields}},
  \href{https://doi.org/10.1016/j.physletb.2015.02.063}{\emph{Phys. Lett. B}
  {\bfseries 743} (2015) 310}
  [\href{https://arxiv.org/abs/1502.06023}{{\ttfamily 1502.06023}}].

\bibitem{Fischer:2023eww}
H.~Fischer, C.~K\"ading, H.~Lemmel, S.~Sponar and M.~Pitschmann, \emph{{Search
  for dark energy with neutron interferometry}},
  \href{https://arxiv.org/abs/2310.18109}{{\ttfamily 2310.18109}}.

\bibitem{Brax:2022olf}
P.~Brax, A.-C.~Davis and B.~Elder, \emph{{Screened scalar fields in hydrogen
  and muonium}}, \href{https://doi.org/10.1103/PhysRevD.107.044008}{\emph{Phys.
  Rev. D} {\bfseries 107} (2023) 044008}
  [\href{https://arxiv.org/abs/2207.11633}{{\ttfamily 2207.11633}}].

\bibitem{Fischer:2023koa}
H.~Fischer, C.~K\"ading, R.I.P.~Sedmik, H.~Abele, P.~Brax and M.~Pitschmann,
  \emph{{Search for environment-dependent dilatons}},
  \href{https://doi.org/10.1016/j.dark.2024.101419}{\emph{Phys. Dark Univ.}
  {\bfseries 43} (2024) 101419}
  [\href{https://arxiv.org/abs/2307.00243}{{\ttfamily 2307.00243}}].

\bibitem{H.Fischer}
H.~Fischer, ``Mathematicacode, version 2.0.0.'' GitHub, Zenodo.
  \url{https://doi.org/10.5281/zenodo.10491446}, 2023.

\bibitem{Fujii2003}
Y.~Fujii and K.-i.~Maeda, \emph{The Scalar-Tensor Theory of Gravitation},
  Cambridge Monographs on Mathematical Physics, Cambridge University Press
  (2003),
  \href{https://doi.org/10.1017/CBO9780511535093}{10.1017/CBO9780511535093}.

\bibitem{pitschmann2023high}
M.~Pitschmann, \emph{The high precision frontier: Search for new physics with
  “tabletop experiments” \& beyond. habilitation}, {\emph{TU Wien} (2023)
  }.

\bibitem{Brax:2022uyh}
P.~Brax, H.~Fischer, C.~K\"ading and M.~Pitschmann, \emph{{The environment
  dependent dilaton in the laboratory and the solar system}},
  \href{https://doi.org/10.1140/epjc/s10052-022-10905-w}{\emph{Eur. Phys. J. C}
  {\bfseries 82} (2022) 934}
  [\href{https://arxiv.org/abs/2203.12512}{{\ttfamily 2203.12512}}].

\bibitem{Corless:1996zz}
R.M.~Corless, G.H.~Gonnet, D.E.G.~Hare, D.J.~Jeffrey and D.E.~Knuth, \emph{{On
  the LambertW function}}, \href{https://doi.org/10.1007/BF02124750}{\emph{Adv.
  Comput. Math.} {\bfseries 5} (1996) 329}.

\bibitem{langtangen2019introduction}
H.P.~Langtangen and K.-A.~Mardal, \emph{Introduction to numerical methods for
  variational problems}, vol.~21, Springer Nature (2019).

\bibitem{leveque2007finite}
R.J.~LeVeque, \emph{Finite difference methods for ordinary and partial
  differential equations: steady-state and time-dependent problems}, SIAM
  (2007).

\bibitem{brune2011exponential}
P.R.~Brune, M.G.~Knepley and L.R.~Scott, \emph{Exponential grids in
  high-dimensional space}, .

\bibitem{Ivanov:2016rfs}
A.N.~Ivanov, G.~Cronenberg, R.~H\"ollwieser, M.~Pitschmann, T.~Jenke,
  M.~Wellenzohn et~al., \emph{{Exact solution for chameleon field, self-coupled
  through the Ratra-Peebles potential with $n=1$ and confined between two
  parallel plates}},
  \href{https://doi.org/10.1103/PhysRevD.94.085005}{\emph{Phys. Rev. D}
  {\bfseries 94} (2016) 085005}
  [\href{https://arxiv.org/abs/1606.06867}{{\ttfamily 1606.06867}}].

\bibitem{tan1990self}
I.-H.~Tan, G.L.~Snider, L.~Chang and E.L.~Hu, \emph{A self-consistent solution
  of schr{\"o}dinger--poisson equations using a nonuniform mesh},
  {\emph{Journal of applied physics} {\bfseries 68} (1990) 4071}.

\bibitem{MarioHabil}
M.~Pitschmann, \emph{{The High Precision Frontier: Search for New Physics with
  “Tabletop Experiments” \& Beyond}}, habilitation, TU Wien, 2023.

\bibitem{olivier2010airy}
V.~Olivier et~al., \emph{Airy functions and applications to physics}, World
  Scientific (2010).

\bibitem{landau1991quantenmechanik}
L.D.~Landau and E.M.~Lifshitz, \emph{Quantenmechanik, lehrbuch der
  theoretischen physik iii},  1991.

\bibitem{nordtvedt2001lunar}
K.~Nordtvedt, \emph{Lunar laser ranging—a comprehensive probe of the
  post-newtonian long range interaction},  in \emph{Gyros, Clocks,
  Interferometers...: Testing Relativistic Graviy in Space}, pp.~317--329,
  Springer (2001).

\bibitem{press2007numerical}
W.H.~Press, \emph{Numerical recipes 3rd edition: The art of scientific
  computing}, Cambridge university press (2007).

\bibitem{mohseni2000numerical}
K.~Mohseni and T.~Colonius, \emph{Numerical treatment of polar coordinate
  singularities}, {\emph{Journal of Computational Physics} {\bfseries 157}
  (2000) 787}.

\end{thebibliography}\endgroup
 
\end{document}